\documentclass[journal=jctcce,manuscript=letter,layout=traditional]{achemso} 

\usepackage[T1]{fontenc} % Use modern font encodings
\usepackage[version=3]{mhchem} 
\usepackage[dvipsnames]{xcolor}
\usepackage{graphicx,subcaption}
\usepackage{amssymb}
\usepackage{overpic}
\usepackage{braket}
\usepackage{comment}
\usepackage{soul}
\usepackage{bm}
\usepackage{hyperref}
\hypersetup{pdfborderstyle={/S/U/W 0.5}}
\usepackage{booktabs}% 
\usepackage{enumitem}
\usepackage{threeparttablex}
\newcommand{\ketbra}[2]{\ensuremath{\left|#1\right\rangle\left\langle#2\right|}}

\title[]{Interference‑Limited Absorption in Dense Molecular Nanolayers Near Reflecting Surfaces}
\author{Zeyu Zhou}
\email{zz9884@princeton.edu}
\affiliation{Department of Chemistry, Princeton University, Princeton, New Jersey 08544, United States}
\author{Maxim Sukharev}
\affiliation{Department of Physics, Arizona State University, Tempe, Arizona 85287, United States}
\alsoaffiliation{College of Integrative Sciences and Arts, Arizona State University, Mesa, Arizona 85212, United States}
\author{Abraham Nitzan}
\affiliation{Department of Chemistry, University of Pennsylvania, 231 South 34th Street, Philadelphia, Pennsylvania 19104, United States}
\alsoaffiliation{School of Chemistry, The Raymond and Beverly Sackler Faculty of Exact Sciences and The Sackler Center for computational Molecular and Materials Science, Tel Aviv University, Tel Aviv 6997801, Israel}
\author{Joseph E. Subotnik}
\email{subotnik@princeton.edu}
\affiliation{Department of Chemistry, Princeton University, Princeton, New Jersey 08544, United States}
\begin{document}
\maketitle

\begin{abstract}
We investigate linear resonant absorption by a dense ensemble of molecules confined to a subwavelength layer in two geometries: (i) a free-standing film in homogeneous space and (ii) the same film placed at a controlled distance from a reflecting surface. In both cases, increasing the effective light–matter coupling (via molecular density/oscillator strength) produces a non-monotonic response: absorption rises to an optimum and then decreases as the film becomes increasingly radiatively bright and reflective. Finite-difference time-domain simulations and analytical transfer-matrix calculations agree quantitatively and yield compact ridge conditions for the optimum. We interpret the trends using a scattering/port picture: the isolated film is a symmetric two-port system (reflection and transmission), which bounds single-sided resonant absorption to $\leq 50\%$ in the ultrathin limit (reflecting transition saturation), whereas adding a mirror suppresses transmission and converts the structure into an effectively one-port absorber. In the mirror-backed geometry, interference can cancel reflection and unity absorption is obtained at critical coupling, when radiative leakage is balanced by intrinsic molecular loss. These results clarify fundamental limits and design rules for collective absorption in dense molecular layers near dielectric or metallic boundaries.
\end{abstract}
\maketitle

\newpage

There is today  a push to design nano-optical devices with  strong optical responses to be used as sensors or for energy harvesting. An important means to control and/or enhance light-matter interactions is to work with molecules located near metallic nanolayers and nanoparticles and/or dielectric mirrors. 
One example is surface enhanced Raman scattering (SERS) where molecular Raman signals can be strongly enhanced near suitable metallic structures.\cite{haynes2005surface, stiles2008surface, frontiera2011surface,sharma2012sers, yi2025surface} 
Another is molecular cavity electrodynamics, where hybridlight-matter states are formed inside the nanostructure.\cite{bayer2001coupling,  walther2006cavity, hutchison2012modifying, ebbesen2016hybrid, kavokin2017microcavities, ribeiro2018polariton, frisk2019ultrastrong, hertzog2019strong, gomez2019energy, garcia2021manipulating, cui2023comparing} 
Overall, understanding how molecules interact with dielectricinhomogeneous environments remains a fruitful topic of investigation. 

To model such physics, there are two effective approaches. 
On the one hand, if one can reasonably capture the behavior of a collection of molecules by introducing a dielectric constant that characterizes the molecular environment, then one can use classical dielectric continuum electromagnetic theory,
to propagate light through just about any material (metallic mirrors, dielectric materials, etc).
For such calculations in simple planar geometries, one can routinely calculate transmission and reflection using a transfer matrix method (TMM). \cite{yeh1990optical}
On the other hand, for nano-devices, if molecule dynamics are important, we must resort to a numerical quantum or semiclassical approach.  
For the latter simulations, it is standard today to run finite-difference time-domain (FDTD) calculations in the presence of dielectric materials, combined with a semiclassical Ehrenfest mean-field description of the matter subsystem.
Several recent works have now demonstrated that a host of interesting strong light matter phenomena can in fact be captured via such mixed quantum-classical algorithms.\cite{taflove2005computational, teixeira2007fdtd, mcmahon2007tailoring, zhao2008methods, sukharev2011numerical, puthumpally2014dipole, sukharev2017optics, you2019nonlinear, sidler2020polaritonic, tancogne2020octopus, chuang2021universal, li2022energy, zhou2024nature}
FDTD together with the Ehrenfest approximation are able to describe a host of optical response phenomena of such structures, and when applied to planar interfaces, to capture the physics revealed by the TMM. 

In the present paper, we use  both TMM and FDTD/Ehrenfest to explore the behavior of a subwavelength layer of molecules (or quantum emitters) at high density, where our goal is to analyze exactly how the optical properties of such a layer differ from the properties of a single molecule (emitter).  We show that, unlike the case of an isolated molecule, the absorption of such a layer of molecules need not have a monotonic behavior on number density; namely, putting more molecules in the sample need not lead to more absorption. 
Most importantly, while we analyze such effects in vacuum, we also investigate such optical behavior near a metallic mirror, where we the non-monotonic effect described above is strongly exacerbated. Overall, our results highlight the strong need to study light-matter interactions from a material (rather than molecular) perspective as several new and unexpected phenomena can be predicted to arise from interference in the high density limit.
The central trend reported here, namely absorption that increases with molecular oscillator strength/density and then decreases, can be understood within the established framework of impedance matching (critical coupling) in planar resonant structures \cite{landy2008perfect, watts2012metamaterial, thongrattanasiri2012complete, kats2013nanometre, ra2013total}. A subwavelength resonant molecular layer behaves as an effective “absorbing sheet” whose response controls both dissipation and re-radiation. In a free-standing (no-mirror) geometry the layer is a two-port scatterer (reflection and transmission), which imposes a well-known constraint: under single sided illumination the resonant absorption of a symmetric ultrathin sheet is bounded (approaching $50\%$ in the thin-layer limit).\cite{chong2010coherent, wan2011time} Introducing a reflecting boundary suppresses transmission and turns the system into an effectively one-port absorber; at appropriate layer–mirror separations the reflected field can be canceled by interference, enabling unity absorption when radiative leakage is balanced by material loss, i.e., at the critical coupling strength. \cite{suh2004temporal} This perspective provides a standard physical interpretation of our ridge conditions and explains why increasing the collective susceptibility beyond an optimum drives the structure into an increasingly reflective regime and reduces absorption.

\subparagraph{Structure of the systems}
\begin{figure}[!ht]
    \includegraphics[width=14cm]{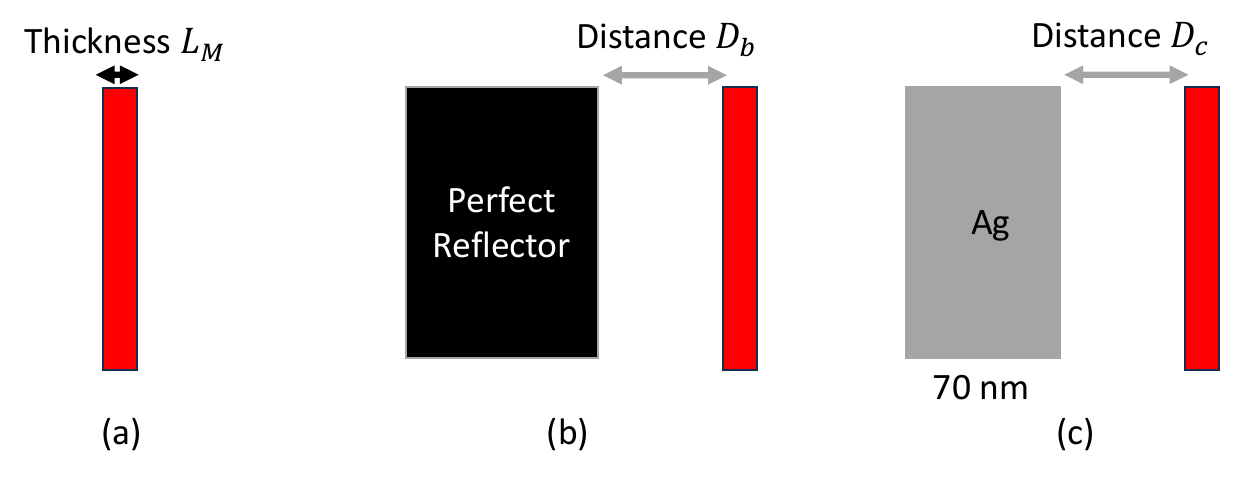}
    \caption{Sketches of the three model scenarios that are studied in this work. (a) A single layer of 2-level molecules with resonant frequency $\hbar\omega_{ge} = 1.9$ eV and thickness $L_{M}$; (b) a single layer of 2-level molecules with near a perfect reflection boundary condition (c) a single layer of 2-level molecules near a thick silver mirror. Both scenarios are probed by white light source ($1$ fs long) from the right-hand side of the system.}
    \label{fig: systemsketch}
\end{figure}
In this work, we consider three scenarios probed by a short ($1$ fs) white pulse.
For the first and simplest scenario, as shown in Fig \ref{fig: systemsketch}(a), we place a single layer of 2-level $\{\ket{g}, \ket{e}\}$ systems with energy spacing $E_{ge} = 1.9$ eV (wavelength $\lambda_{M} = 652$) and transition dipole moment $\mu_{ge} = 5$ Debye. 
Without a mirror, we note that all three signals -- (i) absorption (ii) transmission and (iii) reflection -- will be nonzero in general.
Second, as shown in Fig. \ref{fig: systemsketch} (b), we place a perfect reflection boundary condition (black) near the same 2-level molecular layer (red). 
Here, the perfect reflection boundary condition does not absorb or allow for transmission of the EM field. 
Thus, only (i) molecular absorption (A) and (ii) reflection (R) spectra will be nonzero; they must  add to unity ($T = 0$, $R + A = 1$). 
The goal of this scenario is to study the aforementioned interference effects without any non-perfect mirror effects (absorption and phase shift).

Last, as shown in Fig. \ref{fig: systemsketch} (c), we replace the perfect mirror (black) with a thick ($70$ nm) silver mirror (gray). This scenario (c) replicates a realistic system that can be experimentally probed. Note that in scenarios (b) and (c), we choose the two distances between the mirror and molecular layers to be different ($D_{b} \neq D_{c}$) for the two different dielectric constants of the mirrors; these distances will be determined below by enforcing maximal absorption.

\subparagraph{Maxwell-Bloch equations\label{subsec: maxwellequations}}
The interaction between the external probe is modeled in one-dimension. The one-dimensional Maxwell equations are
\begin{align}
    \frac{\partial {\cal B}_{y}}{\partial t} &= -\frac{\partial {\cal E}_{x}}{\partial{z}}\label{eq: mag}\\
    \epsilon_{0}\frac{\partial {\cal E}_{x}}{\partial t} &= -\frac{\partial {\cal B}_{y}}{\mu_{0}\partial{z}} - J_{x}\label{eq: elec}
\end{align}
where, $\epsilon_{0}$ is the vacuum permittivity, $\mu_{0}$ is the vacuum permeability, $z$ is the longitudinal coordinate that is perpendicular to the mirror and molecular slab and $J_{x}$ is the polarization current,
\begin{align}
    J_{x} = \frac{dP_{x}}{dt} = \varrho\frac{d(\text{Tr}(\hat{\rho}\hat{\mu}_{x}))}{dt}
    \label{eq: current}
\end{align}
Here, $\varrho$ is the number density, $\hat{\rho}$ and $\hat{\mu}_{x}$ are the density matrix and transition dipole operator of the molecular system (see below).

Next, the perfect mirror in scenario (b) is achieved by setting the EM field to be 0 at each time step at the boundary. 
Finally, within the metallic silver mirror, a macroscopic polarization $P_{x}$  is simulated with the Drude model
\begin{align}
\frac{\partial^2 P_{x}}{\partial t^2} + \gamma \frac{\partial P_{x}}{\partial t} = \epsilon_{0}\Omega_{p}^{2}{\cal E}_{x}
\label{eq: silver}
\end{align}
Here, the damping rate $\gamma$ and plasma frequency $\Omega_p$ are taken to fit the dielectric response of silver at optical frequencies. \cite{yang2015optical}
Eqs \ref{eq: mag}-\ref{eq: silver} are discretized in space and time following the finite-difference time-domain (FDTD) approach.
For more details regarding the FDTD solver for Maxwell equations with different dielectric constants, please see \citenum{taflove2005computational}.
Note that the length of the $70$ nm silver mirror is very close to the perfect mirror, so our results are not expected to be too different. 

The molecular layers are described by their $2\times 2$ density matrices $\hat{\rho}$ under a mean-field approximation. In other words, there is one density matrix at each spatial grid point within the molecular layer. The Hamiltonian (and the corresponding Liouvillian) are time-dependent because of the local time-dependent electric field. The system is appended by a phenominological damping with $T_{1} = 1$ ps to ensure numerical stability. Thus, the equation of motion is
\begin{align}
    i\hbar\frac{d}{dt}\hat{\rho}^{j} &= [\hat{H}^{j}(t), \hat{\rho}^{j}] - \frac{i\hbar}
    {T_{1}}
    \left[
\begin{array}{c c}
-\rho^{j}_{ee} & 1/2 \rho^{j}_{ge}\\ 
1/2\rho^{j}_{eg} & \rho^{j}_{ee} \\ 
\end{array} \right]
\end{align}
Here, $j = 1, 2, ..., N$ labels the density matrices at $N$ spatial grid points within the molecular layer.
The EM field enters the Hamiltonian as the coupling between adjacent electronic states, i.e. $\bra{g}\hat{H}^{j}\ket{e} = \mu_{x}^{ge}{\cal E}_{x}^{j}(t)$:
\begin{align}
\hat{H}^{j} =& 0\ketbra{g}{g}+\hbar\omega_{ge}\ketbra{e}{e} + \mu_{x} {\cal E}_{x}^{j}(\ketbra{g}{e} + \ketbra{e}{g}) 
\end{align}
All molecular layers are initialized on the ground state ($\ket{g}$, respectively). \cite{footnote0} Near the edge of the simulation cell, we drive one grid point for $1$ fs as a short pulse.
The spatial grid in our simulations is a one-dimensional array from $-1000$ nm to $1000$ nm, with $4000$ grid points.  The molecular structure in the three scenarios in Fig \ref{fig: systemsketch} are centered at the origin. The laser source is very far from the system (to the right) and generates an EM field traveling in both directions. In the leftward direction, the generated EM field reaches the molecular layer  at normal incidence and $z_{inc} = 950$ nm.
In the rightward direction, the generated EM field is absorbed by a convolutional perfectly matched layer  (CPML) to ensure no reflections as a boundary condition. 
\cite{taflove2005computational}
For all details of parameters, please see appendix \ref{apdx: para}. 

At each time step, we record the EM fields on the two sides of the system: the transmission and reflection signals. These signals are then Fourier transformed and normalized by the incident amplitudes to obtain the transmission ($\mathbf{T}$) and reflection ($\mathbf{R}$) spectra (Obviously $\mathbf{T} = 0$ for the perfectly reflecting mirror). Finally, the absorption spectra are obtained by $\mathbf{A} = 1 -\mathbf{T} - \mathbf{R}$ and exhibit one single peak at $\omega = \omega_{ge} = 1.9 eV$ in all cases. In all figures below, we extract the resonant absorption strength (at $\omega = \omega_{ge} = 1.9$ eV) and vary other parameters of the systems.
\subparagraph{Dielectric constant of a 2-level system}
The absorption spectra can also be obtained by the transfer matrix method, provided that we know the correspondance between a two-level system and its complex-valued dielectric constant. Using the Lorentz model described in ref \citenum{sukharev2017optics}, the dielectric constant is written in the form:
\begin{align}
\chi_{e}(\omega) =& \frac{2\mu_{x}^2\varrho\omega_{ge}}{\varepsilon_{0}(\omega_{ge}^2 - \omega^2+i\omega\Gamma)}
\\
n(\omega) =& \sqrt{1+\chi_{e}(\omega)}
\label{eq: n_eq_sqrt1pchi}
\end{align}
Here, $\hbar\Gamma = 2\pi\hbar/T_{1}\approx 4\times 10^{-3} eV$ represents the phenomenological damping.

We focus on the resonant condition, in which the electric susceptibility is purely imaginary:
\begin{align}
\chi_{e}(\omega = \omega_{ge}) =& -\frac{2i\mu_{x}^2\varrho}{\varepsilon_{0}\Gamma}
\label{eq: chi_resonance}
\end{align}
Below, we define $\chi_{M} = \vert\chi_{e}(\omega = \omega_{ge})\vert$ for characterizing effective light-matter interaction. It is determined by the transition dipole moment $\mu$, the number density $\varrho$ and the lifetime ($1/\Gamma$). 
Note that $1+\chi_{e}(\omega)$ is formally complex-valued; when calculating the dielectric constant $n$ in Eq. \ref{eq: n_eq_sqrt1pchi}, we choose the square root with negative imaginary part ($\Im(n) \leq 0$) meaning the molecular layer will be absorptive. 

All results below are obtained by FDTD calculations and are validated with the transfer matrix method. For detailed discussions of the transfer matrix method, see appendix \ref{apdx: A}.
Let us now begin by investigating how the distance between the mirror and molecular layer changes the absorption strength. We also vary the effective electric susceptibility at resonance to connect the absorption strength and physical properties of the molecular layer, such as number density and transition dipole moment strength.
Thereafter, we study the absorption as a function of varying molecular layer thickness and single molecule electric susceptibility, and present an analytical formula (see below Eq. \ref{eq: max_ridge}) for the absorption maxima.

\subparagraph{Absorption intensity at varying mirror-molecular layer distance}
\begin{figure}[t!]
\includegraphics[width=15cm]{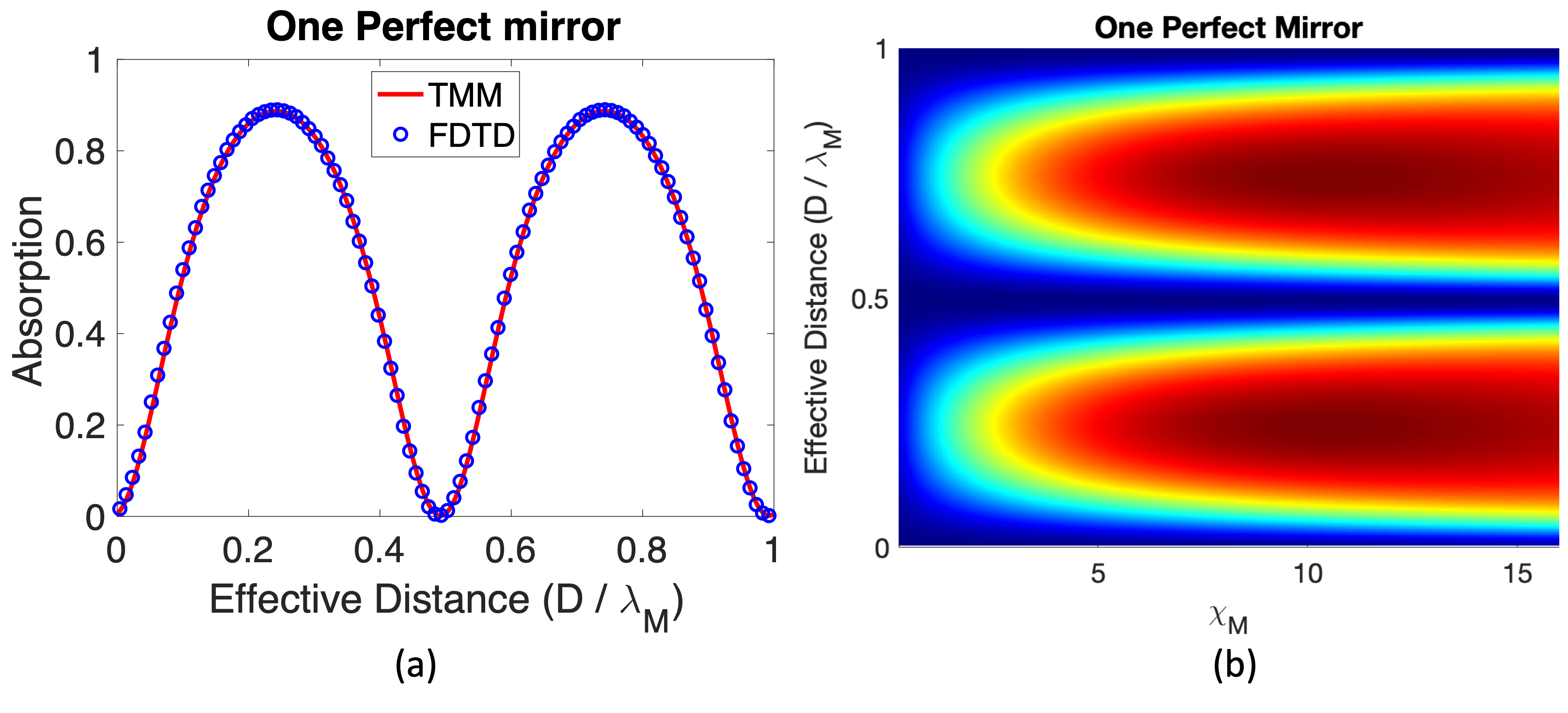}
\caption{(a) Absorption intensity at $\hbar\omega = 1.9$ eV for varying mirror-molecular layer distance. The thickness of the molecular layer is $L_{M} = 10$ nm, the number density is $1$ mol/L, the transition dipole moment is $5$ Debye and these parameter correspond to molecular susceptibility $\chi_{M} = 5.16$. The results are probed by a very weak short pulse and excite minimal population onto the excited state throughout simulation. (b) Absorption intensity plotted against the mirror-molecular layer distance $D$ and the molecular susceptibility $\chi_{M}$ for a perfect reflector. The perfect mirror gives a clear pattern: the maximal absorption values appear at $D = (2N+1)\lambda_{M}/4, N\in\mathbb{N}$ and minimal absorption appears at $D = N\lambda_{M}/2, N\in\mathbb{N}$. This pattern arises due to the constructive/destructive interference of the standing wave.}
\label{fig: distance2d}
\end{figure}

To study the dependence of the absorption on the distance $D$ we use a thin ($L_{M}=10 \text{ nm}\ll D$) molecular layer and calculate the absorption as $D$ is changing within one wavelength. Fig. 2 (a) shows the absorpion calculated using the TMM (red lines) and FDTD (blue lines)
Note that both methods are in agreement here, and the overall prediction is that one finds a peak near $D=(2N+1)\lambda_{M}/4, N \in \mathbb{N}$. 
At this distance, the reflected light undergoes maximal constructive interference with the incident light:
for a perfect mirror, this distance arises because during reflection, the mirror provides an extra $\pi$ phase shift (and thus an effective $\lambda_{M} /2$ light path) which, when combined with another $\lambda_{M} /2$ light path for the round trip between the
molecular layer and the mirror yields a constructive $\lambda_M$ change in phase.
From Fig \ref{fig: distance2d}, we extract the optimal distances (the fundamental interference distance with maximal absorption) $D$ between the molecular layer and the perfect mirror $D_{b} \approx 163$ nm. In and in Appendix \ref{apdx: b}, we have also conducted a similar simulation for a silver mirror, for which we obtain $D_{c} \approx 135$ nm. These distances will be fixed below.

Another aspect of the absorption behavior is seen in Fig. \ref{fig: distance2d} (b), which depicts as a function of both the distance $D$ and the molecular layer susceptibility $\chi_{M}$. The oscillations with distance $D$ reflects the behavior seen in Fig \ref{fig: distance2d} (a). In addition, we observe that the absorption intensity reaches a maximum ($\mathbf{A}(\omega=\omega_{ge}) = 1$, $\mathbf{R}+\mathbf{T} = 0$) at $\chi_{M} \approx 10.5$. Note that the calculation is conducted with a short input white light pulse.
Finally we note that while considering absorption is a standard way to study linear optical response, in the present context transmission and reflection are the natural accessible observables. The corresponding reflection peak (see Appendix \ref{apdx: A}, eqs \ref{eq: abs_reflec_vac} and \ref{eq: abs_reflec_mir}) show the same physical behavior as discussed here. 

\begin{figure}[t!]
\includegraphics[width=\textwidth]{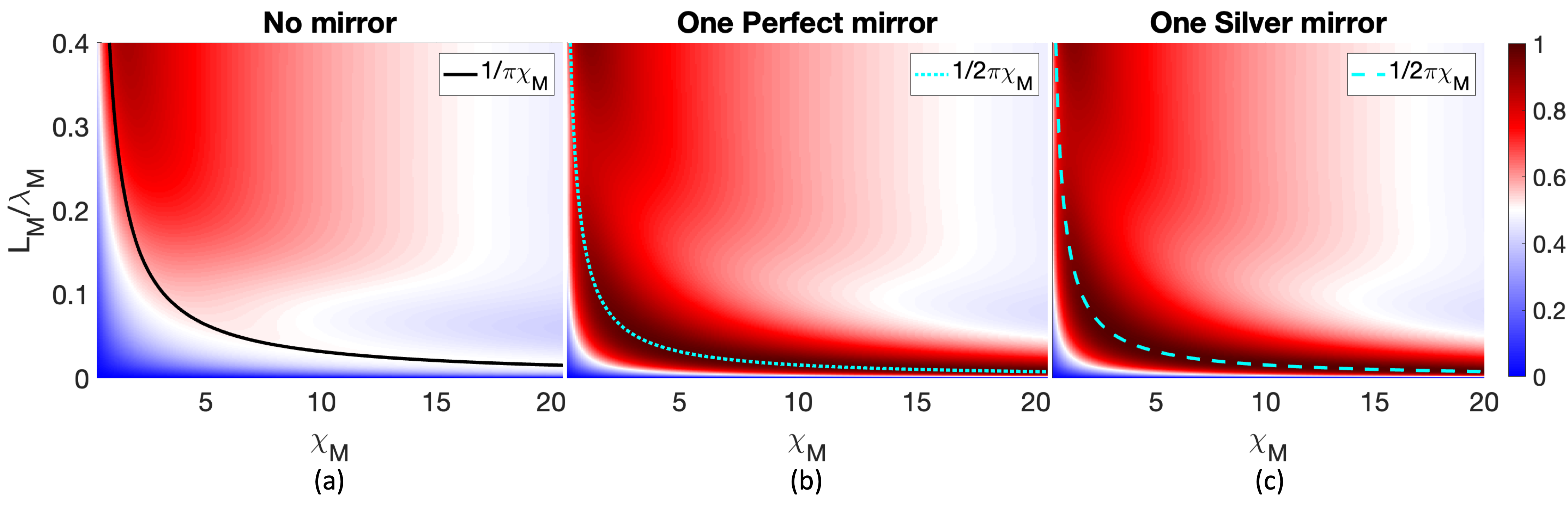}
\caption{Maximal absorption at $\hbar\omega=\hbar\omega_{ge} = 1.9$ eV for varying molecular layer thickness and electric susceptibility of scenario (a) No mirror, (b) one perfect mirror at distance ($D = 163$ nm) and (c) one $70$ nm silver mirror at distance ($D = 135$ nm). In fig (a), because the transmission channel is also finite in scenario (a) and thus by symmetry, the boundary is defined as $A = 0.5$ (white band) and is at $L_M/\lambda_{M} = 1/\pi\chi_{M}$. In contrast, as shown in fig (b) and (c), the maximal absorption ($A = 1$, dark red band) is reached at $L_M/\lambda_{M} = 1/2\pi\chi_{M}$ (cyan lines). In all figures, the lines are most accurate in the limit of $L_{M}/\lambda_{M} < 0.05$.}
\label{fig: thickness}
\end{figure}

\subparagraph{Resonant Absorption Intensity for varying molecular layer thickness \label{subsec: thickness}}
The data above leaves a puzzling question: why is the absorption peaked as as a function of $\chi_{M}$, the light-matter coupling? How is this result intimately tied to the existence of the mirror?
To best answer the questions above, we will next probe the absorption intensity  with no mirror nearby. We allow the molecular layer thickness to change as a free parameter ($L_M$) on the vertical axis in Fig \ref{fig: thickness}.  All data from Fig. \ref{fig: thickness} is generate from a TMM method.

According to Fig. \ref{fig: thickness}(a), we find 
two regimes in the resonant absorption intensity: $L_{M} < 0.05\lambda_{M}$ vs $L_{M} \geq 0.05 \lambda_{M}$. As shown in Fig \ref{fig: thickness} (a), in the  limit of a thin layer thickness $L_{M} < 0.05\lambda_{M}$ for the no mirror scenario (a), we still find maximum in absorption, but now at $A = 0.5$ (white band, highlighted by the black solid line); although perhaps faint to the eye, the absorption on both sides of the ridge is smaller than a half, $A \leq 0.5$ (blue areas).
Interestingly, several of these features change 
 in the opposite limit of large thickness ($L_{M}$).  Here, as one might normally expect, we can find a scenario where $A=1$ (full absorption). Such a  peak (dark red) occurs at $\chi_{M} \approx 2$; for larger  values of $\chi_{M}$, the absorption intensity is reduced  (presumably as the molecular layer becomes more like a mirror).

The data in Fig \ref{fig: thickness} (a) is quite illuminating, insofar as it offers new insight into the different regimes possible. To that end, we next extract the observables for  scenarios (b) and (c), as shown in Fig \ref{fig: thickness} (b) and (c). Now, in the limit of thin layer thickness  $L_{M} < 0.05\lambda_{M}$, the maximal absorption is at $A = 1$ (dark red band, highlighted by cyan dotted and dashed lines) rather than $A = 0.5$ . As already demonstrated in Fig \ref{fig: distance2d}, the nearby mirror enhances the resonant absorption and thus the blue (low absorption) regions near the ridge are narrower than those in the no mirror scenario (a). Finally, as the thickness of the molecular layer $L_{M}$ increases, the resonant absorption is overall stronger than in the scenario without mirror, while the absorption pattern is similar.

\begin{figure}[!t]
    \includegraphics[width=14cm]{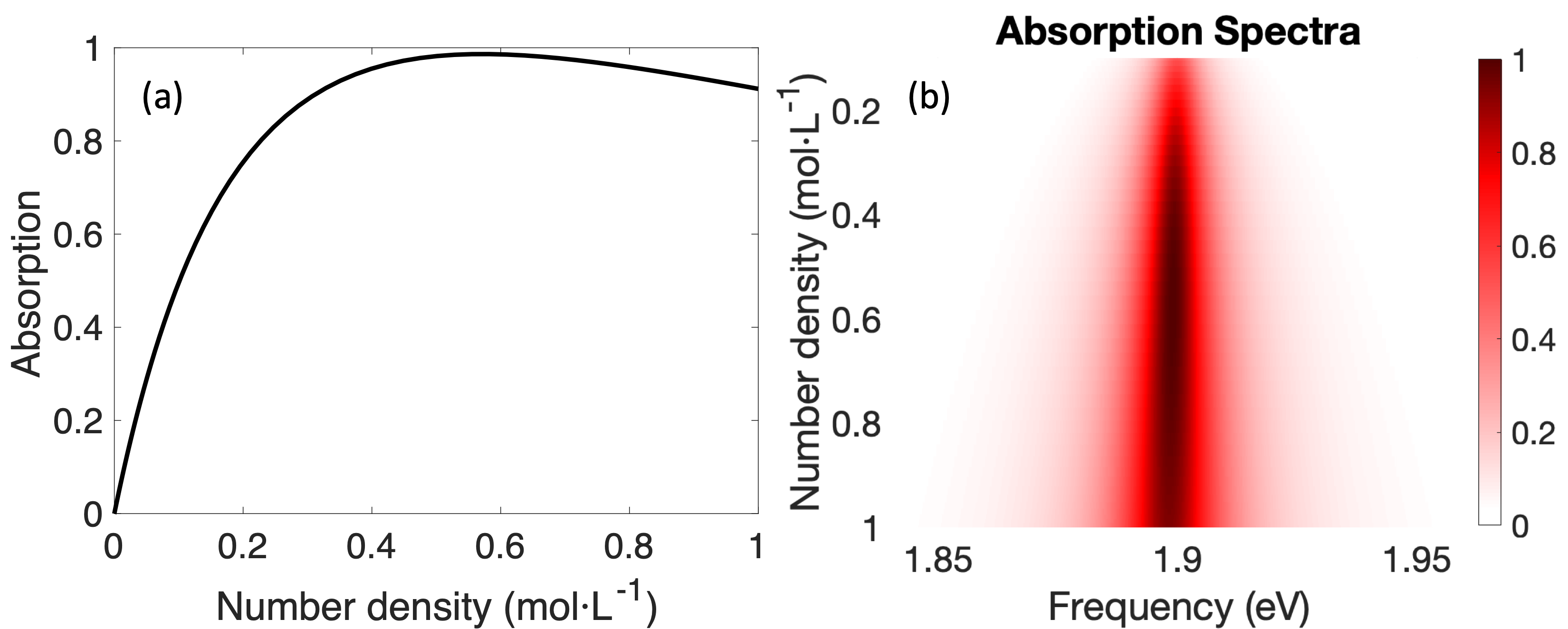}
    \caption{Non-monotonic absorption spectra as a function of the number density.}
    \label{fig: abs_spectra}
\end{figure}
Overall, in comparing Fig \ref{fig: thickness} (a) with (b)(c), the major differences are in the limit $L_{M}\ll \lambda_{M}$, where the absorption peak is larger ($A = 1$ vs $A= 0.5$) and stronger with a mirror as compared to without a mirror. To best understand these effects, we have initiated a series of TMM calculations. As shown in the Appendix \ref{apdx: A}, one can analytically show that indeed, for a narrow molecular layer, the maximal absorption condition becomes $\chi_{M}=k_{M}L_{M}$ which yields $A = 1$ in the presence of a mirror. From Eq. \ref{eq: chi_resonance}, this expression can be written as:
\begin{eqnarray}
    \varrho \mu_x^2 = \frac{\varepsilon_{0}\pi\Gamma L_{M}}{\lambda_{M}}
    \label{eq: max_ridge}
\end{eqnarray}
Without a mirror, the corresponding condition becomes $\chi_{M}=k_{M}L_{M} / 2$ which yields $A = 0.5$.

The fact that the results above can be indeed recovered by a simple TMM calculation (with a simple form for the optimal number density in terms of lengths and wavelengths) indicates that there must be a reasonable simple intuitive explanation. In short, it would appear that for a single molecular layer, a majority of light will pass through the molecular layer for once and leave. Therefore, as the molecular susceptibility increases, the reflection signal increases from $0$ to $1$ while the transmission signal decrease from $1$ to $0$.  By symmetry, one would expect the maximal absorption to be a half (i.e. the other half stays in the form of light and moves away). In contrast, for a molecule near a mirror, the transmission is completely blocked and light can pass through the molecular layer more than once. Hence, the light can move back and forth between the mirror and molecular layer, and larger numbers of wavefronts would seem to allow for more complete interference and stronger maxima.  In any event, one big conclusion is that we need to be careful with using single molecular intuitions when considering a sub-wavelength molecular nanolayer because this nanolayer can exhibit unusual collective responses.

To best reinforce this caveat, in Fig. \ref{fig: abs_spectra}, we plot the absorption on resonance for a two-level system as a function of the number density $\varrho$ (see Eq. \ref{eq: chi_resonance}). Note that there is a maximum, highlighting that more molecular density need not lead to more absorption.  Alternatively, if we focus on the low number density limit, the curve clearly violates Lambert's law: as the number density increases from $0$, the absorption gradually deviates from a linear relation.
However, note that even as we increase the number density, every two-level system remains barely excited and very far from saturated; again the excitation here is collective and cannot be understood in terms of single molecules.

Our analytic conditions can be interpreted as explicit matching conditions between radiative and dissipative channels. In the no-mirror case, the molecular nanolayer necessarily partitions incident power among reflection, transmission, and absorption; increasing the collective susceptibility strengthens both absorption and coherent re-radiation, and beyond a certain point the enhanced back scattering prevents efficient energy delivery into the layer. Consequently, the absorption exhibits a maximum consistent with the thin film two-port bound (approaching 
$A_{max}\approx0.5$ for a symmetric ultrathin resonant sheet under single sided excitation), rather than growing monotonically with oscillator strength or density.

With a mirror present, transmission is eliminated and absorption is controlled primarily by the cancellation (or enhancement) of reflection. At mirror-layer separations that place the molecular sheet near a field antinode (approximately quarter-wave, modified by the mirror reflection phase), the structure behaves as a mirror-backed absorbing sheet analogous to a Salisbury-screen-type absorber. In this one-port configuration, the derived optimal susceptibility corresponds to critical coupling: radiative leakage into the single external channel is matched by internal molecular loss, producing a reflection minimum and (at the optimum) perfect absorption. When the collective susceptibility exceeds this point, the system becomes over-coupled/over-reflective and absorption decreases (providing a unified explanation for the observed non-monotonic dependence of resonant absorption on coupling strength).

In summary, we have investigated the enhancement of resonant absorption spectra of a subwavelength molecular nanolayer by a nearby metallic surface. 
The resonant absorption intensities for varying mirror-molecular layer distance, molecular layer thickness and effective light-matter interactions are obtained by transfer matrix method (TMM) and finite difference time domain (FDTD) calculations. 
We have found that the presence of a mirror enhances the absorption intensity in three aspects: (i)Distance dependent absorption affected by interference between incoming and reflected light with absorption maxima at distances  approximately $(2N+1) \lambda_{M} / 4$, where constructive interferences maximizes the local EM field; 
(ii) there is an optimal pair of light-matter interaction and thickness for maximizing absorption that is different from a bare molecular layer; and (iii) a mirror can make the maximal point both more pronounced (a full absorption) and easier to reach (less collective molecular susceptibility $\chi_{M}$).
Most importantly, for a given choice of molecular layer thickness, as we increase effective light-matter coupling, we observe stronger absorption until a maximum point is reached. However, if we further increase the light-matter coupling, the absorption intensity decreases (which must be due to a convoluted phase alignment between multiple wavefronts) according to both simulation and analytic considerations (see Appendix \ref{apdx: A}).
The root cause is a complex interference effect of many wavefronts; effectively, a molecular layer with its own dielectric must itself act like a cavity to some extent, which automatically leads to a maximum condition for absorption determined by interference; when a mirror is placed, that condition is further enhanced. 

Looking forward,  for a properly chosen set of target molecules, we believe the effect above should be measurable, where we predict a violation of Lambert's law and an enhancement effect due to a mirror.  (Of course, a quantitative interpretation will require properly accounting for  intermolecular coupling \cite{chuang2021universal} when treating a real material. ) The resulting analytic “ridge” conditions provide practical design rules for tuning between reflection dominated, transparency-like, and perfect-absorption regimes in dense resonant films, offering a compact route to engineer narrowband spectral features and strong field–matter interaction in planar nanophotonic architectures. Finally, this work highlights that collective responses from a dense molecular layer can be drastically different from standard intuitions based on single-molecular layer.
In particular, in the nanoregime -- with dense molecular layers situated between molecular and bulk length scales -- the resonance condition in Eq. \ref{eq: max_ridge} offers us an important guide to designing  materials
and nano-devices with unique responses and high sensitivity.

\section*{Acknowledgments}
This work has been supported by the European Research Council (ERC-2024-SyG-101167294; UnMySt) (AN). 
MS research is supported by the Office of Naval Research, Grant No. N000142512090 and the Air Force Office of Scientific Research under Grant No. FA9550-25-1-0096.
This work was also supported by the U.S. Department of Energy, Office of Science, Office of Basic Energy Sciences, under Award No. DOE-SC0025393 (JES)

\appendix

\section{Deriving the maximum boundaries with transfer matrix method \label{apdx: A}}
As shown in Fig \ref{fig: thickness}, the two maximal absorption boundaries are critical to understand the full absorption diagram. Thus, in this appendix, we analytically derive these boundaries with the transfer matrix method (TMM). We will start by introducing the transfer matrix method.

For layered systems, the central physical quantity is the real or complex-valued dielectric constant ($n$). Thereby, for normal incidence onto a layer with thickness $L$, we define matrices:
\begin{align}
\mathbf{D}=\left[
\begin{array}{c c}
1 & 1\\ 
n & -n \\ 
\end{array} \right]
\end{align}
\begin{align}
\mathbf{D}^{-1}=\frac{\left[
\begin{array}{c c}
1 & 1/n\\ 
1 & -1/n \\ 
\end{array} \right]}{2}
\end{align}
\begin{align}
\mathbf{P}(L)=\left[
\begin{array}{c c}
\exp(inkL) & 0\\ 
0 & \exp(-inkL) \\ 
\end{array} \right]
\end{align}
Here, $\mathbf{D}$ describes the interface and $\mathbf{P}$ describes the propagation inside layer. $k$ is the wavevector of the incident light.
The full transfer matrix $\mathbf{M}$ across such a layer is
\begin{align}
\mathbf{M}=\mathbf{DP}(L)\mathbf{D}^{-1} = \frac{\left[
\begin{array}{c c}
\exp(inkL) + \exp(-inkL) & \bigl(\exp(inkL) - \exp(-inkL)\bigr) / n\\ 
n\bigl(\exp(inkL) - \exp(-inkL)\bigr) & \exp(inkL) + \exp(-inkL) \\ 
\end{array} \right]}{2}
\end{align}
Here, $\mathbf{M}$ carries information of how the phase of the EM field with wavevector $k$ evolve after passing through the whole layer and we can extract transmission and reflection from $\mathbf{M}$. For more details of TMM, there is a comprehensive discussion in Ref \citenum{yeh1990optical}.  
Note that we do not reduce the matrix elements to trigonometric functions because in general, the dielectric constant $n$ is complex-valued.
Below, we will separately derive the two boundary lines: No mirror (Fig \ref{fig: thickness} (a)) vs One mirror (Fig \ref{fig: thickness} (b)).

\subsection{No mirror}
The full transfer matrix for any dielectric layer with thickness $L$ is
\footnotesize{\begin{align}
\mathbf{M}=&\mathbf{D}^{-1}_{vac}\mathbf{D}\mathbf{P}(L)\mathbf{D}^{-1}\mathbf{D}_{vac}
\\
=&\frac{\left[
\begin{array}{c c}
1 & 1\\ 
1 & -1 \\ 
\end{array} \right]
\left[
\begin{array}{c c}
\exp(inkL) + \exp(-inkL) & \bigl(\exp(inkL) - \exp(-inkL)\bigr) / n\\ 
n\bigl(\exp(inkL) - \exp(-inkL)\bigr) & \exp(inkL) + \exp(-inkL) \\ 
\end{array} \right]
\left[
\begin{array}{c c}
1 & 1\\ 
1 & -1\\ 
\end{array} \right]}{4}
\end{align}}
Here, we focus on the first column of the matrix ($\mathbf{M}_{11}$ and $\mathbf{M}_{21}$) :
\begin{align}
\mathbf{M}_{11} =& \frac{(n+2+1/n)\exp(inkL) + (2-n-1/n)\exp(-inkL)}{4}
\\
\mathbf{M}_{21} =& \frac{(1/n-n)\exp(inkL) + (n-1/n)\exp(-inkL)}{4}
\end{align}
Next, we calculate the absolute square of the two matrix elements:
\begin{align}
16\vert\mathbf{M}_{11}\vert^2=&\vert n+2+1/n\vert^{2}\exp(-2\Im(n)kL) + \vert 2-n-1/n\vert^{2}\exp(2\Im(n)kL)
\nonumber\\
&+(n+2+1/n)(2-n^{*}-1/n^{*})\exp(2i\Re(n)kL)
\nonumber\\
&+(n^{*}+2+1/n^{*})(2-n-1/n)\exp(-2i\Re(n)kL)
\\
8\vert\mathbf{M}_{11}\vert^2=&\Bigl(4 + \vert n\vert^{2} + \frac{1 + 2\Re(n)^2-2\Im(n)^2}{\vert n\vert^{2}}\Bigr)\cosh(2\Im(n)kL)
\nonumber\\
&-\Bigl(4 \Re(n) + \frac{4 \Re(n)}{\vert n\vert^{2}}\Bigr)\sinh(2\Im(n)kL)
\nonumber\\
&+\Bigl(4 - \vert n\vert^{2} - \frac{1 + 2\Re(n)^2-2\Im(n)^2}{\vert n\vert^{2}}\Bigr)\cos(2\Re(n)kL)
\nonumber\\
&+\Bigl(4 \Im(n) - \frac{4 \Im(n)}{\vert n\vert^{2}}\Bigr)\sin(2\Re(n)kL)
\label{eq: apdx_m11_nomirror}
\end{align}
\begin{align}
8\vert\mathbf{M}_{21}\vert^2=&\Bigl\vert n - \frac{1}{n}\Bigr\vert^2\Bigl(\cosh(2\Im(n)kL)-\cos(2\Re(n)kL)\Bigr)
\nonumber\\
=&(\vert n\vert^{2} + \frac{1 - 2\Re(n)^2+2\Im(n)^2}{\vert n\vert^{2}})\Bigl(\cosh(2\Im(n)kL)-\cos(2\Re(n)kL)\Bigr)
\label{eq: apdx_m21_nomirror}
\end{align}
Here, $\Re(n)$ is the real part and $\Im(n)$ is the imaginary part of the dielectric constant $n$ and thus $\vert n\vert^{2} = \Re(n)^2+\Im(n)^2$. 
As shown in Fig \ref{fig: thickness}, we focus on dense molecules/strong transition dipole moment regime. Thus, the boundary line is most accurate for  small thickness $L$ and large electric susceptibility $\chi_{M}$ regime (lower right). In this limit, we make the following approximation:
\begin{align}
n_{M} =& \sqrt{1+\chi_{e}} \approx \sqrt{-i\chi_{M}}\equiv(1-i)x
\nonumber
\\
x \approx& \Re(n) \approx -\Im(n)\geq0. 
\label{eq: apdx_approx_x}
\\
2kL \equiv& a\geq0
\label{eq: apdx_approx_a}
\end{align}
Thus, the two matrix elements in Eqs \ref{eq: apdx_m11_nomirror} and \ref{eq: apdx_m21_nomirror} become
\begin{align}
\vert\mathbf{M}_{11}\vert^2\approx&\Bigl(\frac{8x^4+4x^2+1}{16x^2}\Bigr)\cosh(ax)+\Bigl(\frac{2x^2+1}{4x}\Bigr)\sinh(ax) + \Bigl(\frac{8x^4-4x^2-1}{16x^2}\Bigr)\cos(ax)+\Bigl(\frac{2x^2-1}{4x}\Bigr)\sin(ax)
\nonumber
\\
\approx&\frac{8x^4+4x^2+1}{16x^2}\Bigl(1+\frac{a^2x^2}{2}\Bigr)+\frac{2ax^2+a}{4x} + \Bigl(\frac{8x^4-4x^2-1}{16x^2}\Bigr)\Bigl(1-\frac{a^2x^2}{2}\Bigr)+\frac{2ax^2-a}{4}
\nonumber
\\
=&\Bigl(\frac{ax^2}{2}+1\Bigr)+\frac{a^2}{16}
\end{align}
\begin{align}
\vert\mathbf{M}_{21}\vert^2\approx&\Bigl(\frac{x^{2}}{4}+\frac{1}{16x^2}\Bigr)\bigl(\cosh(ax) - \cos(ax)\bigr)
\approx \Bigl(\frac{a^2x^{4}}{4}+\frac{a^2}{16}\Bigr)
\end{align}
The final absorption is obtained by
\begin{align}
\mathbf{A} = 1 - \mathbf{T} - \mathbf{R} = \frac{\vert\mathbf{M}_{11}\vert^2 - 1 - \vert\mathbf{M}_{21}\vert^2}{\vert\mathbf{M}_{11}\vert^2}=\frac{4ax^2}{(ax^2+2)^2+\frac{a^2}{4}}\equiv\frac{4x'}{(x'+2)^2+\frac{a^2}{4}}
\label{eq: abs_reflec_vac}
\end{align}
Here, $x' = ax^2$.
Let us calculate the derivative with respect to $x'$ and find the maximum:
\begin{align}
\frac{\partial \mathbf{A}}{\partial x'} =&\frac{a^2-4x'^2+16}{((x'+2)^2+\frac{a^2}{4})^2}=0
\\
\implies 2\pi L_{M}\chi_{M}/\lambda_{M}=ax^2=&x'=\sqrt{4+a^2/4}\approx2
\end{align}
Thus, we obtain an approximate analytical formula for the ridge in the limit of  small thickness for Fig \ref{fig: thickness}(a): $\pi L_{M}\chi_{M}/\lambda_{M} = 1$. The maximal value is 
\begin{align}
\mathbf{A}(x' \approx 2) = \frac{8}{16+\frac{a^2}{4}} \approx \frac{1}{2}
\end{align}
Note that this approximate formula breaks down in the limit of large thickness ($L_{M} / \lambda_{M} > 0.05$), as shown in Fig \ref{fig: thickness}.
\subsection{One Perfect Mirror}
The transfer matrix $\mathbf{D}(\cal M)$ for a perfect mirror is
\begin{align}
\mathbf{D}(\cal M) = \left[
\begin{array}{c c}
1 & 1\\ 
\cal M & -\cal M \\ 
\end{array} \right]
\end{align}
Here, for a perfect mirror, $\cal M$ is a big number so that below  we will omit all terms without $\cal M$.
The full transfer matrix for the  system  in Fig. \ref{fig: systemsketch}(b) is 
\footnotesize{\begin{align}
&\mathbf{M}' = \mathbf{D}^{-1}_{vac}\mathbf{D}\mathbf{P}(L)\mathbf{D}^{-1}\mathbf{D}_{vac}\mathbf{P}_{vac}(D)\mathbf{D}^{-1}_{vac}\mathbf{D}(\cal M)
\nonumber\\\nonumber
=&\frac{\left[
\begin{array}{c c}
1 & 1\\ 
1 & -1 \\ 
\end{array} \right]
\left[
\begin{array}{c c}
\exp(inkL) + \exp(-inkL) & \bigl(\exp(inkL) - \exp(-inkL)\bigr) / n\\ 
n\bigl(\exp(inkL) - \exp(-inkL)\bigr) & \exp(inkL) + \exp(-inkL) \\ 
\end{array} \right]
\left[
\begin{array}{c c}
\cos(kD) & i\sin(kD)\\ 
i\sin(kD)& \cos(kD)\\ 
\end{array} \right]
\left[
\begin{array}{c c}
1 & 1\\ 
\cal M & -\cal M\\ 
\end{array} \right]}{2}
\\
\approx&\frac{\left[
\begin{array}{c c}
1 & 1\\ 
1 & -1 \\ 
\end{array} \right]
\left[
\begin{array}{c c}
\exp(inkL) + \exp(-inkL) & \bigl(\exp(inkL) - \exp(-inkL)\bigr) / n\\ 
n\bigl(\exp(inkL) - \exp(-inkL)\bigr) & \exp(inkL) + \exp(-inkL) \\ 
\end{array} \right]
\left[
\begin{array}{c c}
i{\cal M}\sin(kD) & -i{\cal M}\sin(kD)\\ 
{\cal M}\cos(kD)& -{\cal M}\cos(kD)\\ 
\end{array} \right]
}{2}
\end{align}}
Again, we focus on the two elements of the first column:
\begin{align}
\mathbf{M}'_{11} =& \frac{\cal M}{2}\Bigl\{(n+1)\exp(inkL)\Bigl(i\sin(kD)+\frac{\cos(kD)}{n}\Bigr)+(1-n)\exp(-inkL)\Bigl(i\sin(kD) - \frac{\cos(kD)}{n}\Bigr)\Bigr\}
\\
\mathbf{M}'_{21} =& \frac{\cal M}{2}\Bigl\{(1-n)\exp(inkL)\Bigl(i\sin(kD)+\frac{\cos(kD)}{n}\Bigr)+(n+1)\exp(-inkL)\Bigl(i\sin(kD) - \frac{\cos(kD)}{n}\Bigr)\Bigr\}
\end{align}
Next, we take the absolute square of the two elements:
\begin{align}
\frac{4\vert\mathbf{M}'_{11}\vert^2}{{\cal M}^2} =& \vert 1+n\vert^2\Bigl(\sin^{2}(kD) + \frac{\cos^{2}(kD) + 2\Im(n)\sin(kD)\cos(kD)}{\vert n\vert^2}\Bigr)\exp(-2\Im(n)kL)
\\\nonumber
&+\vert 1-n\vert^2\Bigl(\sin^{2}(kD) + \frac{\cos^{2}(kD) - 2\Im(n)\sin(kD)\cos(kD)}{\vert n\vert^2}\Bigr)\exp(2\Im(n)kL)
\\\nonumber
&+(1-\vert n \vert^2 + 2i\Im(n))\Bigl(\sin^{2}(kD) - \frac{\cos^{2}(kD) - 2i\Re(n)\sin(kD)\cos(kD)}{\vert n\vert^2}\Bigr)\exp(2i\Re(n)kL)
\\\nonumber
&+(1-\vert n \vert^2 - 2i\Im(n))\Bigl(\sin^{2}(kD) - \frac{\cos^{2}(kD) + 2i\Re(n)\sin(kD)\cos(kD)}{\vert n\vert^2}\Bigr)\exp(-2i\Re(n)kL)
\end{align}
\begin{align}
\frac{4\vert\mathbf{M}'_{21}\vert^2}{{\cal M}^2} =& \vert 1-n\vert^2\Bigl(\sin^{2}(kD) + \frac{\cos^{2}(kD) + 2\Im(n)\sin(kD)\cos(kD)}{\vert n\vert^2}\Bigr)\exp(-2\Im(n)kL)
\\\nonumber
&+\vert 1+n\vert^2\Bigl(\sin^{2}(kD) + \frac{\cos^{2}(kD) - 2\Im(n)\sin(kD)\cos(kD)}{\vert n\vert^2}\Bigr)\exp(2\Im(n)kL)
\\\nonumber
&+(1-\vert n \vert^2 - 2i\Im(n))\Bigl(\sin^{2}(kD) - \frac{\cos^{2}(kD) - 2i\Re(n)\sin(kD)\cos(kD)}{\vert n\vert^2}\Bigr)\exp(2i\Re(n)kL)
\\\nonumber
&+(1-\vert n \vert^2 + 2i\Im(n))\Bigl(\sin^{2}(kD) - \frac{\cos^{2}(kD) + 2i\Re(n)\sin(kD)\cos(kD)}{\vert n\vert^2}\Bigr)\exp(-2i\Re(n)kL)
\end{align}
We make the same approximation as eqs \ref{eq: apdx_approx_x} and \ref{eq: apdx_approx_a}:
\begin{align}
\frac{4\vert\mathbf{M}'_{11}\vert^2}{{\cal M}^2} \approx& (1+2x^2+2x)\Bigl(\sin^{2}(kD) + \frac{\cos^{2}(kD) - 2x\sin(kD)\cos(kD)}{2x^2}\Bigr)\exp(ax)
\nonumber\\
&+(1+2x^2-2x)\Bigl(\sin^{2}(kD) + \frac{\cos^{2}(kD) + 2x\sin(kD)\cos(kD)}{2x^2}\Bigr)\exp(-ax)
\nonumber\\
&+(1-2x^2 - 2ix)\Bigl(\sin^{2}(kD) - \frac{\cos^{2}(kD) - 2ix\sin(kD)\cos(kD)}{2x^2}\Bigr)\exp(iax)
\nonumber\\
&+(1-2x^2 + 2ix)\Bigl(\sin^{2}(kD) - \frac{\cos^{2}(kD) + 2ix\sin(kD)\cos(kD)}{2x^2}\Bigr)\exp(-iax)
\nonumber\\
=&\Bigl\{ 2(1+2x^2)\Bigl(\sin^{2}(kD) + \frac{\cos^{2}(kD)}{2x^2}\Bigr)-4\sin(kD)\cos(kD)\Bigr\}\cosh(ax)
\nonumber\\
&+\Bigl\{ 4x\Bigl(\sin^{2}(kD) + \frac{\cos^{2}(kD)}{2x^2}\Bigr)-\frac{2(1+2x^2)}{x}\sin(kD)\cos(kD)\Bigr\}\sinh(ax)
\nonumber\\
&+\Bigl\{ 2(1-2x^2)\Bigl(\sin^{2}(kD) - \frac{\cos^{2}(kD)}{2x^2}\Bigr)+4\sin(kD)\cos(kD)\Bigr\}\cos(ax)
\nonumber\\
&+\Bigl\{ 4x\Bigl(\sin^{2}(kD) - \frac{\cos^{2}(kD)}{2x^2}\Bigr)-\frac{2(1-2x^2)}{x}\sin(kD)\cos(kD)\Bigr\}\sin(ax)
\nonumber\\
\approx&\Bigl\{ 2(1+2x^2)\Bigl(\sin^{2}(kD) + \frac{\cos^{2}(kD)}{2x^2}\Bigr)-4\sin(kD)\cos(kD)\Bigr\}(1+a^2x^2/2)
\nonumber\\
&+\Bigl\{ 4x\Bigl(\sin^{2}(kD) + \frac{\cos^{2}(kD)}{2x^2}\Bigr)-\frac{2(1+2x^2)}{x}\sin(kD)\cos(kD)\Bigr\}(ax)
\nonumber\\
&+\Bigl\{ 2(1-2x^2)\Bigl(\sin^{2}(kD) - \frac{\cos^{2}(kD)}{2x^2}\Bigr)+4\sin(kD)\cos(kD)\Bigr\}(1-a^2x^2/2)
\nonumber\\
&+\Bigl\{ 4x\Bigl(\sin^{2}(kD) - \frac{\cos^{2}(kD)}{2x^2}\Bigr)-\frac{2(1-2x^2)}{x}\sin(kD)\cos(kD)\Bigr\}(ax)
\nonumber\\
=&4(ax^2+1)^2\sin^2(kD)+(4 + a^2)\cos^2(kD) - (4+4ax^2)\sin(kD)\cos(kD)
\end{align}
\begin{align}
\frac{4\vert\mathbf{M}'_{21}\vert^2}{{\cal M}^2} \approx&(1+2x^2-2x)\Bigl(\sin^{2}(kD) + \frac{\cos^{2}(kD) - 2x\sin(kD)\cos(kD)}{2x^2}\Bigr)\exp(ax)
\nonumber\\
&+(1+2x^2+2x)\Bigl(\sin^{2}(kD) + \frac{\cos^{2}(kD) + 2x\sin(kD)\cos(kD)}{2x^2}\Bigr)\exp(-ax)
\nonumber\\
&+(1-2x^2 + 2ix)\Bigl(\sin^{2}(kD) - \frac{\cos^{2}(kD) - 2ix\sin(kD)\cos(kD)}{2x^2}\Bigr)\exp(iax)
\nonumber\\
&+(1-2x^2 - 2ix)\Bigl(\sin^{2}(kD) - \frac{\cos^{2}(kD) + 2ix\sin(kD)\cos(kD)}{2x^2}\Bigr)\exp(-iax)
\nonumber\\
=&\Bigl\{ 2(1+2x^2)\Bigl(\sin^{2}(kD) + \frac{\cos^{2}(kD)}{2x^2}\Bigr)+4\sin(kD)\cos(kD)\Bigr\}\cosh(ax)
\nonumber\\
&-\Bigl\{ 4x\Bigl(\sin^{2}(kD) + \frac{\cos^{2}(kD)}{2x^2}\Bigr)+\frac{2(1+2x^2)}{x}\sin(kD)\cos(kD)\Bigr\}\sinh(ax)
\nonumber\\
&+\Bigl\{ 2(1-2x^2)\Bigl(\sin^{2}(kD) - \frac{\cos^{2}(kD)}{2x^2}\Bigr)-4\sin(kD)\cos(kD)\Bigr\}\cos(ax)
\nonumber\\
&-\Bigl\{ 4x\Bigl(\sin^{2}(kD) - \frac{\cos^{2}(kD)}{2x^2}\Bigr)+\frac{2(1-2x^2)}{x}\sin(kD)\cos(kD)\Bigr\}\sin(ax)
\nonumber\\
\approx&4(ax^2-1)^2\sin^2(kD)+(4 + a^2)\cos^2(kD) + (4-4ax^2)\sin(kD)\cos(kD)
\end{align}
Note that the transmission is
\begin{align}
\mathbf{T} = \frac{1}{\vert\mathbf{M}'_{11}\vert^2} \sim \frac{1}{{\cal M}^2}\approx0
\end{align}
Thus finally, the absorption is
\begin{align}
\mathbf{A} \approx 1-\mathbf{R} = &1-\frac{4(ax^2-1)^2\sin^2(kD)+(4 + a^2)\cos^2(kD) + (4-4ax^2)\sin(kD)\cos(kD)}{4(ax^2+1)^2\sin^2(kD)+(4 + a^2)\cos^2(kD) - (4+4ax^2)\sin(kD)\cos(kD)} 
\nonumber\\
=&\frac{16ax^2\sin^2(kD) - 8ax^2\sin(kD)\cos(kD)}{4(ax^2+1)^2\sin^2(kD)+(4 + a^2)\cos^2(kD) - (4+4ax^2)\sin(kD)\cos(kD)}
\label{eq: abs_reflec_mir}
\end{align}
As shown in Fig \ref{fig: distance2d}, we focus on the maximal constructive interference limit: $kD \approx \pi/2$. 
Hence, 
\begin{align}
\mathbf{A} \approx \frac{4ax^2}{(ax^2+1)^2} \equiv\frac{4x'}{(x'+1)^2}
\end{align}
We will find the maximum by taking the derivative of $\mathbf{A}$ with respect to $x'$
\begin{align}
\frac{\partial \mathbf{A}}{\partial x'} =& \frac{4(x'-1)}{(x'+1)^3} = 0
\\
2\pi L_{M}\chi_{M}/\lambda_{M}=ax^2=&x'=1
\end{align}
Hence finally, we obtain the approximate analytical formula for the ridge in the limit of  small thickness for Fig \ref{fig: thickness}(b): $2\pi L_{M}\chi_{M}/\lambda_{M} = 1$. The maximal absorption is $\mathbf{A}(x'=1) = 1$.
\section{A realistic Silver mirror\label{apdx: b}}
\begin{figure}[h!]
\includegraphics[width=\textwidth]{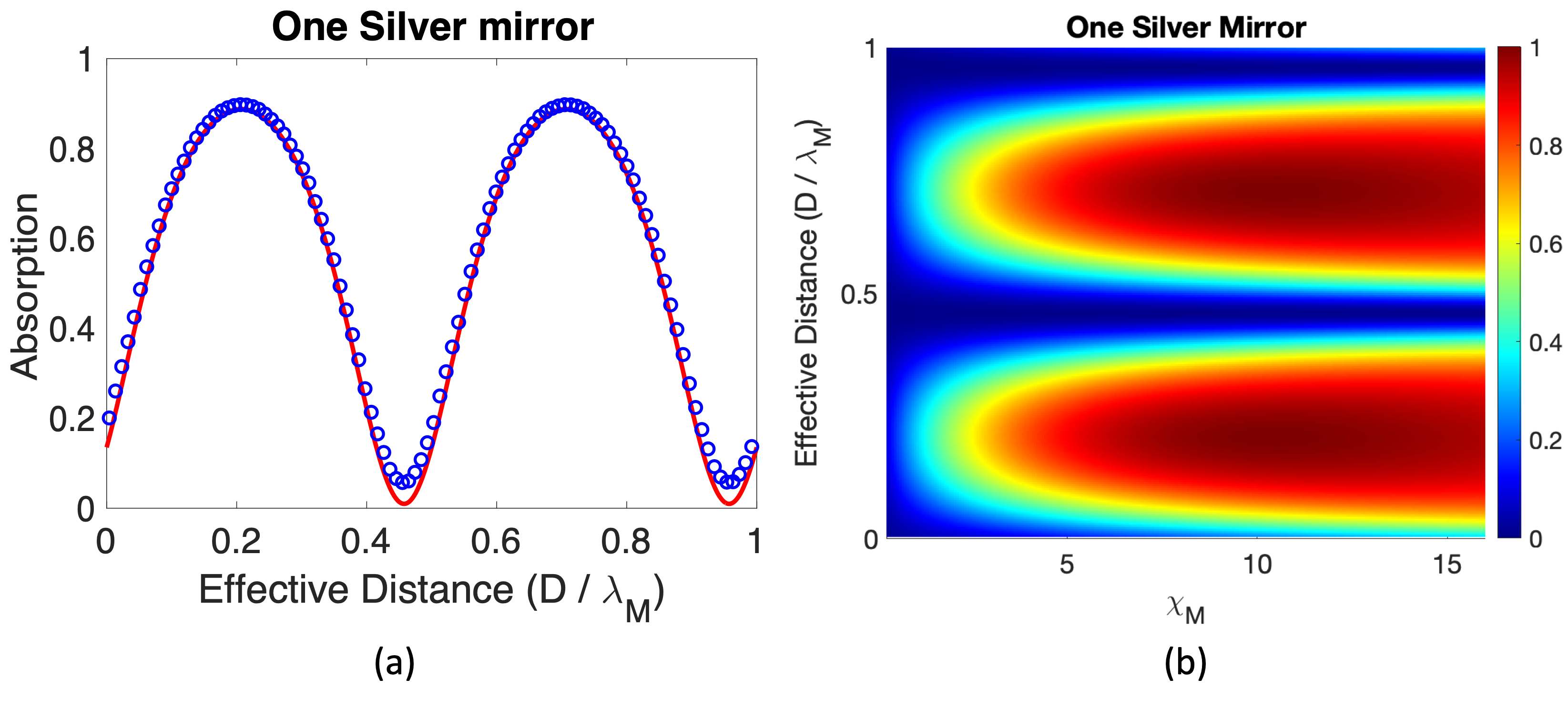}
\caption{Absorption at resonance, $\hbar\omega=\hbar\omega_{ge} = 1.9$ eV plotted against (a) the mirror-molecular layer distance $D$ and (b) the molecular susceptibility $\chi_{M}$ for a silver mirror. The silver mirror possesses a finite complex-valued dielectric constant and thus the dependence of maximal/minimal absorption on effective distance deviates from (a). The thickness of molecular layer is $10$ nm.}
\label{fig: distance}
\end{figure}
In this appendix, we present the supplementary results obtained for the more realistic silver mirror, we first note that the absorption behavior is not very different from that obtained for the ideal mirror. Some interesting differences may still be seen: the fundamental interference distance is smaller than $\lambda_{M}/4$ because a realistic silver mirror with a finite complex-valued dielectric constant $n_{Ag}$ is penetrable and thus we only need a smaller distance $D_{c}$. 
Moreover, a realistic and penetrable silver mirror breaks the quantitative agreement between TMM and FDTD. For perfect agreement, one must also require that within the frequency window of interest, the real-valued dielectric constant must dominate over the complex-valued dielectric constant (e.g., using a dielectric mirror). 
After all, TMM cannot fully capture the absorption due to irreversible loss from the metallic structure and molecular intrinsic relaxation. 
With this in mind, as shown in Fig. \ref{fig: distance}(a), TMM shows an unphysical vanishing of the absorption in the case where all EM signals pass through the system and get reflected by the silver mirror; TMM completely misses all irreversible absorption effects due to penetration depth and thus cannot rigorously agree with FDTD.
\section{Parameters\label{apdx: para}}
In this appendix, we list all parameters in the TMM and FDTD simulations.

\begin{table}[h!]
  \begin{threeparttable}
   \caption[]{Parameters for Maxwell-Bloch simulation.}
   \centering
   \begin{tabular}{cc}
     \midrule 
     Name & Value\tnote{$\dagger$}
    \\\hline
    Molecules resonant frequency $\hbar\omega_{ge}$ & $1.9$ eV
    \\
    Molecules resonant wavelength $\lambda_{M}$ & $652$ nm
    \\
    Thickness of the silver mirror & $70$ nm
    \\
    Distance between silver mirror and molecular layer in Fig \ref{fig: systemsketch}(c) & $130$ nm
    \\
    Drude model for silver & $\Omega_{d} = 8.90$eV
    \\
    & $\Gamma_{d} = 0.243$ eV
    \\
    Transition dipole moment & $\mu_{x}= 5$ Debye
    \\
    Simulation grid size (1D)& $2000$ nm
    \\
    Number of grid points & $4000$
    \\
    Number of density matrices & $20$
    \\
    Grid resolution ($dx$) & $0.5$ nm
    \\
    Time step (dt) & $dx/2/c_{0}$
    \\
    CPML grid points & 19
    \\
    \midrule
     \end{tabular}
  \end{threeparttable}
\end{table}
\section*{REFERENCES}
\bibliography{main}
\end{document}